\documentclass[twocolumn,10pt]{article}

\usepackage{graphics}
\usepackage{times}

\begin{document}

\def\etal{{\it{}et~al.}}        
\def\eref#1{(\protect\ref{#1})}
\def\fref#1{\protect\ref{#1}}
\def\sref#1{\protect\ref{#1}}
\def\tref#1{\protect\ref{#1}}
\def\av#1{\langle#1\rangle}
\def\half{\mbox{$\frac12$}}
\def\O{{\rm O}}
\def\vr{{\bf r}}

\newdimen\narrowfigwidth
\setlength{\narrowfigwidth}{2.3in}
\def\columnfigure#1{\resizebox{\linewidth}{!}{\includegraphics{#1}}}
\def\narrowfigure#1{\resizebox{\narrowfigwidth}{!}{\includegraphics{#1}}}
\def\twocolfigure#1{\resizebox{\textwidth}{!}{\includegraphics{#1}}}

\title{Models of the Small World\\
\large A Review}
\author{M. E. J. Newman\\
{\normalsize\it Santa Fe Institute, 1399 Hyde Park Road, Santa Fe, NM 87501}}
\date{}
\maketitle

\begin{abstract}
  It is believed that almost any pair of people in the world can be
  connected to one another by a short chain of intermediate acquaintances,
  of typical length about six.  This phenomenon, colloquially referred to
  as the ``six degrees of separation,'' has been the subject of
  considerable recent interest within the physics community.  This paper
  provides a short review of the topic.
\end{abstract}

\section{Introduction}
The United Nations Department of Economic and Social Affairs estimates that
the population of the world exceeded six billion people for the first time
on October~12, 1999.  There is no doubt that the world of human society has
become quite large in recent times.  Nonetheless, people routinely claim
that, global statistics notwithstanding, it's still a small world.  And in
a certain sense they are right.  Despite the enormous number of people on
the planet, the structure of social networks---the map of who knows
whom---is such that we are all very closely connected to one another
(Kochen~1989, Watts~1999).

One of the first quantitative studies of the structure of social networks
was performed in the late 1960s by Stanley Milgram, then at Harvard
University (Milgram~1967).  He performed a simple experiment as follows.
He took a number of letters addressed to a stockbroker acquaintance of his
in Boston, Massachusetts, and distributed them to a random selection of
people in Nebraska.  (Evidently, he considered Nebraska to be about as far
as you could get from Boston, in social terms, without falling off the end
of the world.)  His instructions were that the letters were to be sent to
their addressee (the stockbroker) by passing them from person to person,
and that, in addition, they could be passed only to someone whom the passer
new on a first-name basis.  Since it was not likely that the initial
recipients of the letters were on a first-name basis with a Boston
stockbroker, their best strategy was to pass their letter to someone whom
they felt was nearer to the stockbroker in some social sense: perhaps
someone they knew in the financial industry, or a friend in Massachusetts.

A reasonable number of Milgram's letters did eventually reach their
destination, and Milgram found that it had only taken an average of six
steps for a letter to get from Nebraska to Boston.  He concluded, with a
somewhat cavalier disregard for experimental niceties, that six was
therefore the average number of acquaintances separating the pairs of
people involved, and conjectured that a similar separation might
characterize the relationship of any two people in the entire world.  This
situation has been labeled ``six degrees of separation'' (Guare~1990), a
phrase which has since passed into popular folklore.

Given the form of Milgram's experiment, one could be forgiven for supposing
that the figure six is probably not a very accurate one.  The experiment
certainly contained many possible sources of error.  However, the general
result that two randomly chosen human beings can be connected by only a
short chain of intermediate acquaintances has been subsequently verified,
and is now widely accepted (Korte and Milgram~1970).  In the jargon of the
field this result is referred to as the {\bf small-world effect}.

The small-world effect applies to networks other than networks of friends.
Brett Tjaden's parlor game ``The Six Degrees of Kevin Bacon'' connects any
pair of film actors via a chain of at most eight co-stars (Tjaden and
Wasson~1997).  Tom Remes has done the same for baseball players who have
played on the same team (Remes~1997).  With tongue very firmly in cheek,
the {\it New York Times\/} played a similar game with the the names of
those who had tangled with Monica Lewinsky (Kirby and Sahre~1998).

All of this however, seems somewhat frivolous.  Why should a serious
scientist care about the structure of social networks?  The reason is that
such networks are crucially important for communications.  Most human
communication---where the word is used in its broadest sense---takes place
directly between individuals.  The spread of news, rumors, jokes, and
fashions all take place by contact between individuals.  And a rumor can
spread from coast to coast far faster over a social network in which the
average degree of separation is six, than it can over one in which the
average degree is a hundred, or a million.  More importantly still, the
spread of disease also occurs by person-to-person contact, and the
structure of networks of such contacts has a huge impact on the nature of
epidemics.  In a highly connected network, this year's flu---or the HIV
virus---can spread far faster than in a network where the paths between
individuals are relatively long.

In this paper we outline some recent developments in the theory of social
networks, particularly in the characterization and modeling of networks,
and in the modeling of the spread of information or disease.

\section{Random graphs}
\label{random}
The simplest explanation for the small-world effect uses the idea of a
random graph.  Suppose there is some number $N$ of people in the world, and
on average they each have $z$ acquaintances.  This means that there are
$\half Nz$ connections between people in the entire population.  The number
$z$ is called the {\bf coordination number} of the network.

We can make a very simple model of a social network by taking $N$ dots
(``nodes'' or ``vertices'') and drawing $\half Nz$ lines (``edges'')
between randomly chosen pairs to represent these connections.  Such a
network is called a {\bf random graph} (Bollob\'as~1985).  Random graphs
have been studied extensively in the mathematics community, particularly by
Erd\"os and R\'enyi~(1959).  It is easy to see that a random graph shows
the small-world effect.  If a person~A on such a graph has $z$ neighbors,
and each of A's neighbors also has $z$ neighbors, then A has about $z^2$
second neighbors.  Extending this argument A also has $z^3$ third
neighbors, $z^4$ fourth neighbors and so on.  Most people have between a
hundred and a thousand acquaintances, so $z^4$ is already between about
$10^8$ and $10^{12}$, which is comparable with the population of the world.
In general the number $D$ of degrees of separation which we need to
consider in order to reach all $N$ people in the network (also called the
{\bf diameter} of the graph) is given by setting $z^D = N$, which implies
that $D = \log N/\log z$.  This logarithmic increase in the number of
degrees of separation with the size of the network is typical of the
small-world effect.  Since $\log N$ increases only slowly with $N$, it
allows the number of degrees to be quite small even in very large systems.

As an example of this type of behavior, Albert~\etal~(1999) studied the
properties of the network of ``hyperlinks'' between documents on the World
Wide Web.  They estimated that, despite the fact there were
$N\simeq8\times10^8$ documents on the Web at the time the study was carried
out, the average distance between documents was only about~19.

There is a significant problem with the random graph as a model of social
networks however.  The problem is that people's circles of acquaintance
tend to overlap to a great extent.  Your friend's friends are likely also
to be your friends, or to put it another way, two of your friends are
likely also to be friends with one another.  This means that in a real
social network it is not true to say that person~A has $z^2$ second
neighbors, since many of those friends of friends are also themselves
friends of person~A.  This property is called {\bf clustering} of networks.

\begin{table}[t]
\begin{center}
\begin{tabular}{l|c|c|c|c}
Network & $N$ & $\ell$ & $C$ & $C_{\rm rand}$ \\
\hline
movie actors   & $225\,226$ & $3.65$ & $0.79$ & $0.00027$ \\
neural network & $282$      & $2.65$ & $0.28$ & $0.05$ \\
power grid     & $4941$     & $18.7$ & $0.08$ & $0.0005$ \\
\end{tabular}
\end{center}
\caption{The number of nodes $N$, average degree of separation $\ell$, and
  clustering coefficient $C$, for three real-world networks.  The last
  column is the value which $C$ would take in a random graph with the same
  size and coordination number.
\label{wstable}}
\end{table}

A random graph does not show clustering.  In a random graph the probability
that two of person~A's friends will be friends of one another is no greater
than the probability that two randomly chosen people will be.  On the other
hand, clustering {\em has\/} been shown to exist in a number of real-world
networks.  One can define a {\bf clustering coefficient} $C$, which is the
average fraction of pairs of neighbors of a node which are also neighbors
of each other.  In a fully connected network, in which everyone knows
everyone else, $C=1$; in a random graph $C=z/N$, which is very small for a
large network.  In real-world networks it has been found that, while $C$ is
significantly less than~1, it is much greater than $\O(N^{-1})$.  In
Table~\tref{wstable}, we show some values of $C$ calculated by Watts and
Strogatz~(1998) for three different networks: the network of collaborations
between movie actors discussed previously, the neural network of the worm
{\it C.~Elegans,} and the Western Power Grid of the United States.  We also
give the value $C_{\rm rand}$ which the clustering coefficient would have
on random graphs of the same size and coordination number, and in each case
the measured value is significantly higher than for the random graph,
indicating that indeed the graph is clustered.

In the same table we also show the average distance $\ell$ between pairs of
nodes in each of these networks.  This is not the same as the diameter $D$
of the network discussed above, which is the {\it maximum} distance between
nodes, but it also scales at most logarithmically with number of nodes on
random graphs.  This is easy to see, since the average distance is strictly
less than or equal to the maximum distance, and so $\ell$ cannot increase
any faster than $D$.  As the table shows, the value of $\ell$ in each of
the networks considered is small, indicating that the small-world effect is
at work.  (The precise definition of ``small-world effect'' is still a
matter of debate, but in the present case a reasonable definition would be
that $\ell$ should be comparable with the value it would have on the random
graph, which for the systems discussed here it is.)

So, if random graphs do not match well the properties of real-world
networks, is there an alternative model which does?  Such a model has been
suggested by Duncan Watts and Steven Strogatz.  It is described in the next
section.

\begin{figure*}
\twocolfigure{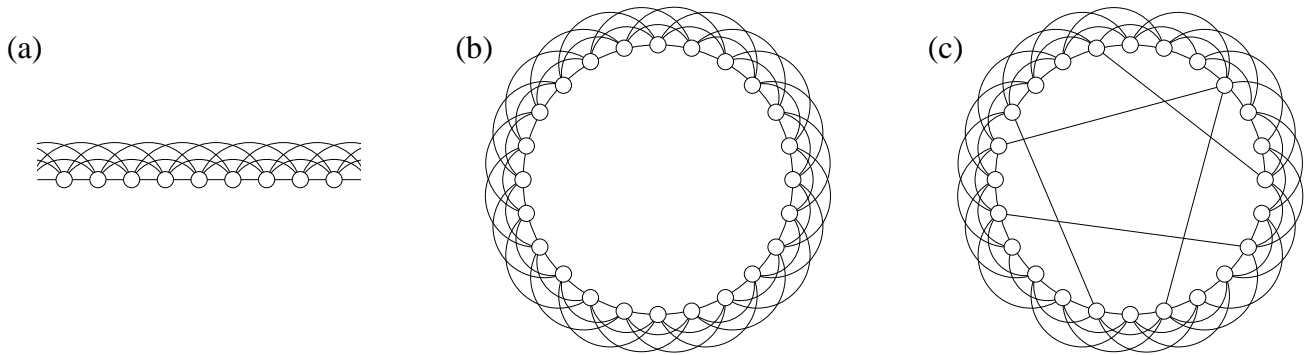}
\caption{(a)~A one-dimensional lattice with each site connected to its $z$
nearest neighbors, where in this case $z=6$.  (b)~The same lattice with
periodic boundary conditions, so that the system becomes a ring.  (c)~The
Watts--Strogatz model is created by rewiring a small fraction of the links
(in this case five of them) to new sites chosen at random.}
\label{model}
\end{figure*}

\section{The small-world model of Watts and Strogatz}
\label{ws}
In order to model the real-world networks described in the last section, we
need to find a way of generating graphs which have both the clustering and
small-world properties.  As we have argued, random graphs show the
small-world effect, possessing average vertex-to-vertex distances which
increase only logarithmically with the total number $N$ of vertices, but
they do not show clustering---the property that two neighbors of a vertex
will often also be neighbors of one another.

The opposite of a random graph, in some sense, is a completely ordered
lattice, the simplest example of which is a one-dimensional lattice---a set
of vertices arranged in a straight line.  If we take such a lattice and
connect each vertex to the $z$ vertices closest to it, as in
Fig.~\fref{model}a, then it is easy to see that most of the immediate
neighbors of any site are also neighbors of one another, i.e.,~it shows the
clustering property.  Normally, we apply periodic boundary conditions to
the lattice, so that it wraps around on itself in a ring
(Fig.~\fref{model}b), although this is just for convenience and not
strictly necessary.  For such a lattice we can calculate the clustering
coefficient $C$ exactly.  As long as $z<\frac23 N$, which it will be for
almost all graphs, we find that
\begin{equation}
C = {3(z-2)\over4(z-1)},
\end{equation}
which tends to $\frac34$ in the limit of large $z$.  We can also build
networks out of higher-dimensional lattices, such as square or cubic
lattices, and these also show the clustering property.  The value of the
clustering coefficient in general dimension $d$ is
\begin{equation}
C = {3(z-2d)\over4(z-d)},
\label{gend}
\end{equation}
which also tends to $\frac34$ for $z\gg2d$.

Low-dimensional regular lattices however do not show the small-world effect
of typical vertex--vertex distances which increase only slowly with system
size.  It is straightforward to show that for a regular lattice in $d$
dimensions which has the shape of a square or (hyper)cube of side $L$, and
therefore has $N=L^d$ vertices, the average vertex--vertex distance
increases as $L$, or equivalently as $N^{1/d}$.  For small values of $d$
this does not give us small-world behavior.  In one dimension for example,
it means that the average distance increases linearly with system size.  If
we allow the dimension $d$ of the lattice to become large, then $N^{1/d}$
becomes a slowly increasing function of $N$, and so the lattice does show
the small-world effect.  Could this be the explanation for what we see in
real networks?  Perhaps real networks are roughly regular lattices of very
high dimension.  This explanation is in fact not unreasonable, although it
has not been widely discussed.  It works quite well, provided the mean
coordination number $z$ of the vertices is much higher than twice the
dimension $d$ of the lattice.  (If we allow $z$ to approach $2d$, then the
clustering coefficient, Eq.~\eref{gend}, tends to zero, implying that the
lattice loses its clustering properties.)

Watts and Strogatz~(1998) however have proposed an alternative model for
the small world, which perhaps fits better with our everyday intuitions
about the nature of social networks.  Their suggestion was to build a model
which is, in essence, a low-dimensional regular lattice---say a
one-dimensional lattice---but which has some degree of randomness in it,
like a random graph, to produce the small-world effect.  They suggested a
specific scheme for doing this as follows.  We take the one-dimensional
lattice of Fig.~\fref{model}b, and we go through each of the links on the
lattice in turn and, with some probability $p$, we randomly ``rewire'' that
link, meaning that we move one of its ends to a new position chosen at
random from the rest of the lattice.  For small $p$ this produces a graph
with is still mostly regular but has a few connections which stretch long
distances across the lattice as in Fig.~\fref{model}c.  The coordination
number of the lattice is still $z$ on average as it was before, although
the number of neighbors of any particular vertex can be greater or smaller
than $z$.

In social terms, we can justify this model by saying that, while most
people are friends with their immediate neighbors---neighbors on the same
street, people that they work with, people that their friends introduce
them to---some people are also friends with one or two people who are a
long way away, in some social sense---people in other countries, people
from other walks of life, acquaintances from previous eras of their lives,
and so forth.  These long-distance acquaintances are represented by the
long-range links in the model of Watts and Strogatz.

Clearly the values of the clustering coefficient $C$ for the
Watts--Strogatz model with small values of $p$ will be close to those for
the perfectly ordered lattice given above, which tend to~$\frac34$ for
fixed small $d$ and large $z$.  Watts and Strogatz also showed by numerical
simulation that the average vertex--vertex distance $\ell$ is comparable
with that for a true random graph, even for quite small values of $p$.  For
example, for a random graph with $N=1000$ and $z=10$, they found that the
average distance was about $\ell=3.2$ between two vertices chosen at
random.  For their rewiring model, the average distance was only slightly
greater, at $\ell=3.6$, when the rewiring probability $p=\frac14$, compared
with $\ell=50$ for the graph with no rewired links at all.  And even for
$p=\frac{1}{64}=0.0156$, they found $\ell=7.4$, a little over twice the
value for the random graph.  Thus the model appears to show both the
clustering and small-world properties simultaneously.  This result has
since been confirmed by further simulation as well as analytic work on
small-world models, which is described in the next section.

\section{Analytic and numerical results for small-world models}
\label{analytic}
Most of the recent work on models of the small world has been performed
using a variation of the Watts--Strogatz model suggested by Newman and
Watts~(1999a).  In this version of the model, instead of rewiring links
between sites as in Fig.~\fref{model}c, extra links, often called {\bf
  shortcuts}, are added between pairs of sites chosen at random, but no
links are removed from the underlying lattice.  This model is somewhat
easier to analyze than the original Watts--Strogatz model, because it is
not possible for any region of the graph to become disconnected from the
rest, whereas this can happen in the original model.  Mathematically a
disjoint section of the graph can be represented by saying that the
distance from any vertex in that section to a vertex somewhere on the rest
of the graph is infinite.  However, this means that, when averaged over all
possible realizations of the graph, the average vertex--vertex distance
$\ell$ in the model is also infinite for any finite value of $p$.  (A
similar problem in the theory of random graphs is commonly dealt with by
averaging the reciprocal of vertex--vertex distance, rather than the
distance itself, but this approach does not seem to have been tried for the
Watts--Strogatz model.)  In fact, it is possible to show that the series
expansion of $\ell/L$ in powers of $p$ about $p=0$ is well-behaved up to
order $p^{z-1}$, but that the expansion coefficients are infinite for all
higher orders.  For the version of the model where no links are ever
removed, the expansion coefficients take the same values up to order
$p^{z-1}$, but are finite for all higher orders as well.  Generically, both
versions of the model are referred to as {\bf small-world models}, or
sometimes {\bf small-world graphs}.

\begin{figure}
\columnfigure{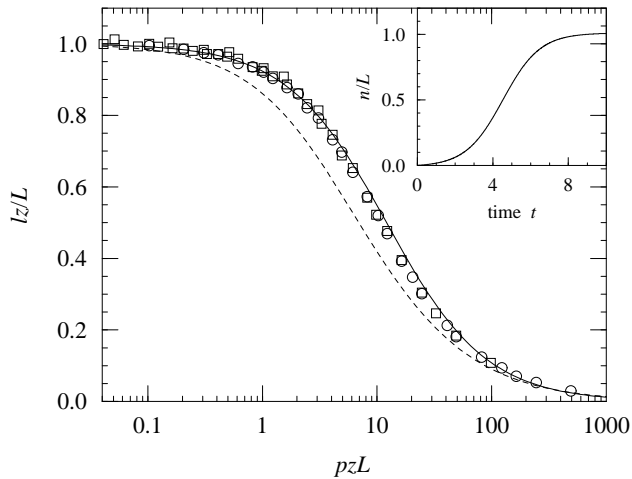}
\caption{Scaling collapse of average vertex--vertex distances on
  $d=1$ small-world graphs according to Eq.~\eref{eqscale}.  Points are
  numerical data for $z=2$ (circles) and $z=10$ (squares), for a variety of
  values of $p$ and $L$.  The solid line is a Pad\'e approximant derived
  from series expansions of the scaling function, while the dotted line is
  the mean-field solution, Eq.~\eref{mft}.  Inset: the number of people
  infected as a function of time by a disease which starts with a single
  person and spreads through a community with the topology of a small-world
  graph.  After Newman and Watts~(1999b) and Newman~\etal~(2000).}
\label{scaling}
\end{figure}

Many results have been derived for small-world models, and many of their
other properties have been explored numerically.  Here we give only a brief
summary of the most important results.  Barth\'el\'emy and Amaral~(1999)
conjectured that the average vertex--vertex distance $\ell$ obeys the
scaling form $\ell = \xi G(L/\xi)$, where $G(x)$ is a universal scaling
function of its argument $x$ and $\xi$ is a characteristic length-scale for
the model which is assumed to diverge in the limit of small $p$ according
to $\xi\sim p^{-\tau}$.  On the basis of numerical results, Barth\'el\'emy
and Amaral further conjectured that $\tau=\frac23$.  Barrat~(1999)
disproved this second conjecture using a simple physical argument which
showed that $\tau$ cannot be less than~1, and suggested on the basis of
more numerical results that in fact it was exactly~1.  Newman and
Watts~(1999b) showed that the small-world model has only one non-trivial
length-scale other than the lattice spacing, which we can equate with the
variable $\xi$ above, and which is given by
\begin{equation}
\xi = {1\over pz}
\end{equation}
for the one-dimensional model, or
\begin{equation}
\xi = {1\over(p z d)^{1/d}}
\end{equation}
in the general case.  Thus $\tau$ must indeed be~1 for $d=1$, or $\tau=1/d$
for general $d$ and, since there are no other length-scales present, $\ell$
must be of the form
\begin{equation}
\ell = {L\over2dz} F(pzL^d),
\label{eqscale}
\end{equation}
where $F(x)$ is another universal scaling function.  (The initial factor of
$(2d)^{-1}$ before the scaling function is arbitrary.  It is chosen thus to
give $F$ a simple limit for small values of its argument---see
Eq.~\eref{limits}.)  This scaling form is equivalent to that of
Barth\'el\'emy and Amaral by the substitution $G(x)=x F(x)$ if $\tau=1$.
It has been extensively confirmed by numerical simulation (Newman and
Watts~1999a, de~Menezes~\etal~2000) and by series expansions (Newman and
Watts~1999b) (see Fig.~\fref{scaling}).  The divergence of $\xi$ as $p\to0$
gives something akin to a critical point in this limit.  (De
Menezes~\etal~(2000) have argued that, for technical reasons, we should
refer to this point as a ``first order critical point'' (Fisher and
Berker~1982).)  This allowed Newman and Watts~(1999a) to apply a real-space
renormalization group transformation to the model in the vicinity of this
point and prove that the scaling form above is exactly obeyed in the limit
of small $p$ and large $L$.

Eq.~\eref{eqscale} tells us that although the average vertex--vertex
distance on a small-world graph appears at first glance to be a function of
three parameters---$p$, $z$, and $L$---it is in fact entirely determined by
a single scalar function of a single scalar variable.  If we know the form
of this one function, then we know everything.  Actually, this statement is
strictly only true if $\xi\gg1$, when it is safe to ignore the other
length-scale in the problem, the lattice parameter of the underlying
lattice.  Thus, the scaling form is expected to hold only when $p$ is
small, i.e.,~in the regime where the majority of a person's contacts are
local and only a small fraction long-range.  (The fourth parameter $d$ also
enters the equation, but is not on an equal footing with the others, since
the functional form of $F$ changes with $d$, and thus Eq.~\eref{eqscale}
does not tell us how $\ell$ varies with dimension.)

Both the scaling function $F(x)$ and the scaling variable $x\equiv pzL^d$
have simple physical interpretations.  The variable $x$ is two times the
average number of shortcuts on the graph for the given value of $p$, and
$F(x)$ is the average fraction by which the vertex--vertex distance on the
graph is reduced for the given value of $x$.  From the results shown in
Fig.~\fref{scaling}, we can see that it takes about $5\frac12$ shortcuts to
reduce the average vertex-vertex distance by a factor of two, and 56 to
reduce it by a factor of ten.

In the limit of large $p$ the small-world model becomes a random graph or
nearly so.  Hence, we expect that the value of $\ell$ should scale
logarithmically with system size $L$ when $p$ is large, and also, as the
scaling form shows, when $L$ is large.  On the other hand, when $p$ or $L$
is small we expect $\ell$ to scale linearly with $L$.  This implies that
$F(x)$ has the limiting forms
\begin{equation}
F(x) = \left\lbrace \begin{array}{ll}
            \mbox{$1$}       & \qquad\mbox{for $x\ll1$}\\
            (\log x)/x         & \qquad\mbox{for $x\gg1$.}
          \end{array} \right.
\label{limits}
\end{equation}
In theory there should be a leading constant in front of the large-$x$ form
here, but, as discussed shortly, it turns out that this constant is equal
to unity.  The cross-over between the small- and large-$x$ regimes must
happen in the vicinity of $L=\xi$, since $\xi$ is the only length-scale
available to dictate this point.

Neither the actual distribution of path lengths in the small-world model
nor the average path length $\ell$ has been calculated exactly yet; exact
analytical calculations have proven very difficult for the model.  Some
exact results have been given by Kulkarni~\etal~(2000) who show, for
example, that the value of $\ell$ is simply related to the mean $\av{s}$
and mean square $\av{s^2}$ of the shortest distance $s$ between two points
on diametrically opposite sides of the graph, according to
\begin{equation}
{\ell\over L} = {\av{s}\over L-1} - {\av{s^2}\over L(L-1)}.
\end{equation}
Unfortunately, calculating the shortest distance between opposite points is
just as difficult as calculating $\ell$ directly, either analytically or
numerically.

Newman~\etal~(2000) have calculated the form of the scaling function $F(x)$
for $d=1$ small-world graphs using a mean-field-like approximation, which
is exact for small or large values of $x$, but not in the regime where
$x\simeq1$.  Their result is
\begin{equation}
F(x) = {4\over\sqrt{x^2+4x}} \tanh^{-1} {x\over\sqrt{x^2+4x}}.
\label{mft}
\end{equation}
This form is also plotted on Fig.~\fref{scaling} (dotted line).  Since this
is exact for large $x$, it can be expanded about $1/x=0$ to show that the
leading constant in the large-$x$ form of $F(x)$, Eq.~\eref{limits}, is~$1$
as stated above.

Newman~\etal\ also solved for the complete distribution of lengths between
vertices in the model within their mean-field approximation.  This
distribution can be used to give a simple model of the spread of a disease
in a small world.  If a disease starts with a single person somewhere in
the world, and spreads first to all the neighbors of that person, and then
to all second neighbors, and so on, then the number of people $n$ who have
the disease after $t$ time-steps is simply the number of people who are
separated from the initial carrier by a distance of $t$ or less.  Newman
and Watts~(1999b) previously gave an approximate differential equation for
$n(t)$ on an infinite small-world graph, which they solved for the
one-dimensional case; Moukarzel~(1999) later solved it for the case of
general~$d$.  The mean-field treatment generalizes the solution for $d=1$
to finite lattice sizes.  (A similar mean-field result has been given for a
slightly different disease-spreading model by Kleczkowski and
Grenfell~(1999).)  The resulting form for $n(t)$ is shown in the inset of
Fig.~\fref{scaling}, and clearly has the right general sigmoidal shape for
the spread of an epidemic.  In fact, this form of $n$ is typical also of
the standard logistic growth models of disease spread, which are mostly
based on random graphs (Sattenspiel and Simon~1988, Kretschmar and
Morris~1996).  In the next section we consider some (slightly) more
sophisticated models of disease spreading on small-world graphs.

\section{Other models based on small-world graphs}
A variety of authors have looked at dynamical systems defined on
small-world graphs built using either the Watts--Strogatz rewiring method
or the alternative method described in Section~\sref{analytic}.  We briefly
describe a number of these studies in this section.

Watts and Strogatz~(1998, Watts~1999) looked at cellular automata, simple
games, and networks of coupled oscillators on small-world networks.  For
example, they found that it was much easier for a cellular automaton to
perform the task known as density classification (Das~\etal~1994) on a
small-world graph than on a regular lattice; they found that in an iterated
multi-player game of Prisoner's Dilemma, cooperation arose less frequently
on a small-world graph than on a regular lattice; and they found that the
small-world topology helped oscillator networks to synchronize much more
easily than in the regular lattice case.

Monasson~(1999) investigated the eigenspectrum of the Laplacian operator on
small-world graphs using a transfer matrix method.  This spectrum tells us
for example what the normal modes would be of a system of masses and
springs built with the topology of a small-world graph.  Or, perhaps more
usefully, it can tell us how diffusive dynamics would occur on a small
world graph; any initial state of a diffusive field can be decomposed into
eigenvectors which each decay independently and exponentially with a decay
constant related to the corresponding eigenvalue.  Diffusive motion might
provide a simple model for the spread of information of some kind in a
social network.

Barrat and Weigt~(2000) have given a solution of the ferromagnetic Ising
model on a $d=1$ small-world network using a replica method.  Since the
Ising model has a lower critical dimension of two, we would expect it not
to show a phase transition when $p=0$ and the graph is truly
one-dimensional.  On the other hand, as soon as $p$ is greater than zero,
the effective dimension of the graph becomes greater than one, and
increases with system size (Newman and Watts~1999b).  Thus for any finite
$p$ we would expect to see a phase transition at some finite temperature in
the large system limit.  Barrat and Weigt confirmed both analytically and
numerically that indeed this is the case.  The Ising model is of course a
highly idealized model, and its solution in this context is, to a large
extent, just an interesting exercise.  However, the similar problem of a
Potts antiferromagnet on a general graph has real practical applications,
e.g.,~in the solution of scheduling problems.  Although this problem has
not been solved on the small-world graph, Walsh~(1999) has found results
which indicate that it may be interesting from a computational complexity
point of view; finding a ground state for a Potts antiferromagnet on a
small-world graph may be significantly harder than finding one on either a
regular lattice or a random graph.

Newman and Watts~(1999b) looked at the problem of disease spread on
small-world graphs.  As a first step away from the very simple models of
disease described in the last section, they considered a disease to which
only a certain fraction $q$ of the population is susceptible; the disease
spreads neighbor to neighbor on a small-world graph, except that it only
affects, and can be transmitted by, the susceptible individuals.  In such a
model, the disease can only spread within the connected cluster of
susceptible individuals in which it first starts, which is small if $q$ is
small, but becomes larger, and eventually infinite, as $q$ increases.  The
point at which it becomes infinite---the point at which an epidemic takes
place---is precisely the percolation point for site percolation with
probability $q$ on the small-world graph.  Newman and Watts gave an
approximate calculation of this epidemic point, which compares reasonably
favorably with their numerical simulations.  Moore and Newman~(2000a,
2000b) later gave an exact solution.

Lago-Fern\'andez~\etal~(2000) investigated the behavior of a neural network
of Hodgkin--Huxley neurons on a variety of graphs, including regular
lattices, random graphs, and small-world graphs.  They found that the
presence of a high degree of clustering in the network allowed the network
to establish coherent oscillation, while short average vertex--vertex
distances allowed the network to produce fast responses to changes in
external stimuli.  The small-world graph, which simultaneously possesses
both of these properties, was the only graph they investigated which showed
both coherence and fast response.

Kulkarni~\etal~(1999) studied numerically the behavior of the Bak--Sneppen
model of species coevolution (Bak and Sneppen~1993) on small-world graphs.
This is a model which mimics the evolutionary effects of interactions
between large numbers of species.  The behavior of the model is known to
depend on the topology of the lattice on which it is situated, and Kulkarni
and co-workers suggested that the topology of the small-world graph might
be closer to that of interactions in real ecosystems than the
low-dimensional regular lattices on which the Bak--Sneppen model is usually
studied.  The principal result of the simulations was that on a small-world
graph the amount of evolutionary activity taking place at any given vertex
varies with the coordination number of the vertex, with the most connected
nodes showing the greatest activity and the least connected ones showing
the smallest.

\section{Other models of the small world}
Although most of the work reviewed in this article is based on the
Watts--Strogatz small-world model, a number of other models of social
networks have been proposed.  In Section~\sref{random} we mentioned the
simple random-graph model and in Section~\sref{ws} we discussed a model
based on a regular lattice of high dimension.  In this section we describe
briefly three others which have been suggested.

\begin{figure}
\begin{center}
\narrowfigure{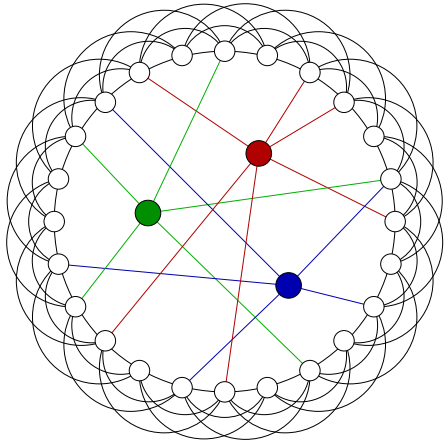}
\end{center}
\caption{An alternative model of a small world, in which there are a small
number of individuals who are connected to many widely-distributed
acquaintances.}
\label{alter}
\end{figure}

One alternative to the view put forward by Watts and Strogatz is that the
small-world phenomenon arises not because there are a few ``long-range''
connections in the otherwise short-range structure of a social network, but
because there are a few nodes in the network which have unusually high
coordination numbers (Kasturirangan~1999) or which are linked to a widely
distributed set of neighbors.  Perhaps the ``six degrees of separation''
effect is due to a few people who are particularly well connected.
(Gladwell~(1998) has written a lengthy and amusing article arguing that a
septuagenarian salon proprietor in Chicago named Lois Weisberg is an
example of precisely such a person.)  A simple model of this kind of
network is depicted in Fig.~\fref{alter}, in which we start again with a
one-dimensional lattice, but instead of adding extra links between pairs of
sites, we add a number of extra vertices in the middle which are connected
to a large number of sites on the main lattice, chosen at random.  (Lois
Weisberg would be one of these extra sites.)  This model is similar to the
Watts--Strogatz model in that the addition of the extra sites effectively
introduces shortcuts between randomly chosen positions on the lattice, so
it should not be surprising to learn that this model does display the
small-world effect.  In fact, even in the case where only one extra site is
added, the model shows the small-world effect if that site is sufficiently
highly connected.  This case has been solved exactly by Dorogovtsev and
Mendes~(1999).

Another alternative model of the small world has been suggested by
Albert~\etal~(1999) who, in their studies of the World Wide Web discussed
in Section~\sref{random}, concluded that the Web is dominated by a small
number of very highly connected sites, as described above, but they also
found that the distribution of the coordination numbers of sites (the
number of ``hyperlinks'' pointing to or from a site) was a power-law,
rather than being bimodal as it is in the previous model.  They produced a
model network of this kind as follows.  Starting with a normal random graph
with average coordination number $z$ and the desired number $N$ of
vertices, they selected a vertex at random and added a link between it and
another randomly chosen site if that addition would bring the overall
distribution of coordination numbers closer to the required power law.
Otherwise the vertex remains as it is.  If this process is repeated for a
sufficiently long time, a network is generated with the correct
coordination numbers, but which is in other respects a random graph.
In particular, it does not show the clustering property of which such a
fuss has been made in the case of the Watts--Strogatz model.  Albert~\etal\ 
found that their model matched the measured properties of the World Wide
Web quite closely, although related work by Adamic~(1999) indicates that
clustering is present in the Web, so that the model is unrealistic in this
respect.

It is worth noting that networks identical to those of Albert~\etal\ can be
generated in a manner much more efficient than the Monte Carlo scheme
described above by simply generating $N$ vertices with a power law
distribution of lines emerging from them (using, for instance, the
transformation method (Newman and Barkema~1999)), and then joining pairs of
lines together at random until none are left.  If one were interested in
investigating such networks numerically, this would probably be the best
way to generate them.

A third suggestion has been put forward by Kleinberg~(1999), who argues
that a model such as that of Watts and Strogatz, in which shortcuts connect
vertices arbitrarily far apart with uniform probability, is a poor
representation of at least some real-world situations.
(Kasturirangan~(1999) has made a similar point.)  Kleinberg notes that in
the real world, people are surprisingly good at finding short paths between
pairs of individuals (Milgram's letter experiment, and the Kevin Bacon game
are good examples) given only local information about the structure of the
network.  Conversely, he has shown that no algorithm exists which is
capable of finding such paths on networks of the Watts--Strogatz type,
again given only local information.  Thus there must be some additional
properties of real-world networks which make it possible to find short
paths with ease.  To investigate this question further, Kleinberg has
proposed a generalization of the Watts--Strogatz model in which the typical
distance traversed by the shortcuts can be tuned.  Kleinberg's model is
based on a two-dimensional square lattice (although it could be generalized
to other dimensions $d$ in a straightforward fashion) and has shortcuts
added between pairs of vertices $i, j$ with probability which falls off as
a power law $d_{ij}^{\,-r}$ of the distance between them.  (In this work,
$d_{ij}$ is the ``Manhattan distance'' $|x_i-x_j|+|y_i-y_j|$, where
$(x_i,y_i)$ and $(x_j,y_j)$ are the lattice coordinates of the vertices $i$
and $j$.  This makes good sense, since this is also the distance in terms
of links on the underlying lattice that separates those two points before
the shortcuts are added.  However, one could in principle generate networks
using a different definition of distance, such as the Euclidean distance
$\sqrt{(x_i-x_j)^2+(y_i-y_j)^2}$, for example.)  It is then shown that for
the particular value $r=2$ of the exponent of the power law (or $r=d$ for
underlying lattices of $d$ dimensions), there exists a simple algorithm for
finding a short path between two given vertices, making use only of local
information.  For any other value of $r$ the problem of finding a short
path is provably much harder.  This result demonstrates that there is more
to the small world effect than simply the existence of short paths.

\section{Conclusions}
In this article we have given an overview of recent theoretical work on the
``small-world'' phenomenon.  We have described in some detail the
considerable body of recent results dealing with the Watts--Strogatz
small-world model and its variants, including analytic and numerical
results about network structure and studies of dynamical systems on
small-world graphs.

What have we learned from these efforts and where is this line of research
going now?  The most important result is that small-world graphs---those
possessing both short average person-to-person distances and ``clustering''
of acquaintances---show behaviors very different from either regular
lattices or random graphs.  Some of the more interesting such behaviors are
the following:
\begin{enumerate}
\setlength{\itemsep}{0pt}
\item These graphs show a transition with increasing number of vertices
  from a ``large-world'' regime in which the average distance between two
  people increases linearly with system size, to a ``small-world'' one in
  which it increases logarithmically.
\item This implies that information or disease spreading on a small-world
  graph reaches a number of people which increases initially as a power of
  time, then changes to an exponential increase, and then flattens off as
  the graph becomes saturated.
\item Disease models which incorporate a measure of susceptibility to
  infection have a percolation transition at which an epidemic sets in,
  whose position is influenced strongly by the small-world nature of the
  network.
\item Dynamical systems such as games or cellular automata show
  quantitatively different behavior on small-world graphs and regular
  lattices.  Some problems, such as density classification, appear to be
  easier to solve on small-world graphs, while others, such as scheduling
  problems, appear to be harder.
\item Some real-world graphs show characteristics in addition to the
  small-world effect which may be important to their function.  An
  example is the World Wide Web, which appears to have a scale-free
  distribution of the coordination numbers of vertices.
\end{enumerate}

Research in this field is continuing in a variety of directions.  Empirical
work to determine the exact structure of real networks is underway in a
number of groups, as well as theoretical work to determine the properties
of the proposed models.  And studies to determine the effects of the
small-world topology on dynamical processes, although in their infancy,
promise an intriguing new perspective on the way the world works.

\section*{Acknowledgements}
The author would like to thank Luis Amaral, Marc Barth\'el\'emy, Rahul
Kulkarni, Cris Moore, Cristian Moukarzel, Naomi Sachs, Steve Strogatz, Toby
Walsh, and Duncan Watts for useful discussions and comments.  This work was
supported in part by the Santa Fe Institute and DARPA under grant number
ONR N00014-95-1-0975.

\def\refer#1#2#3#4#5#6{\item{\frenchspacing\sc#1}\hspace{4pt}
                       #2\hspace{8pt}#3 {\frenchspacing#4} {\bf#5}, #6.}
\def\bookref#1#2#3#4{\item{\frenchspacing\sc#1}\hspace{4pt}
                     #2\hspace{8pt}{\it#3}  #4.}

\section*{References}
$^*$Citations of the form {\tt cond-mat/xxxxxxx} refer to the online
condensed matter physics preprint archive at {\tt http://www.arxiv.org/}.

\baselineskip=15pt
\begin{list}{}{\leftmargin=2em \itemindent=-\leftmargin%
\itemsep=3pt \parsep=0pt \small}
\item {\sc Adamic, L. A.}\ \ 1999\ \ The small world web.  Available
  as {\tt ftp://parcftp.xerox.com/pub/dynamics/\-smallworld.ps}.
\refer{Albert, R., Jeong, H. and Barab\'asi, A.-L.}{1999}{Diameter of the
  world-wide web.}{\it Nature\/}{401}{130--131}
\refer{Bak, P. and Sneppen, K.}{1993}{Punctuated equilibrium and
  criticality in a simple model of evolution.}{\it Physical Review
  Letters}{71}{4083--4086}
\refer{Barth\'el\'emy, M. and Amaral, L. A. N.}{1999}{Small-world networks:
  Evidence for a crossover picture.}{\it Physical Review
  Letters\/}{82}{3180--3183}
\item {\frenchspacing\sc Barrat, A.}\ \ 1999\ \ Comment on ``Small-world
  networks: Evidence for a crossover picture.''  Available as {\tt
  cond-mat/9903323}.
\refer{Barrat, A. and Weigt, M.}{2000}{On the properties of small-world
  network models.}{\it European Physical Journal B\/}{13}{547--560}
\bookref{Bollob\'as, B.}{1985}{\it Random Graphs.}{Academic Press (New
  York)}
\item {\frenchspacing\sc Das, R., Mitchell, M., and Crutchfield, J. P.}\ \
  1994\ \ A genetic algorithm discovers particle-based computation in
  cellular automata.  In {\it Parallel Problem Solving in Nature,} Davidor,
  Y., Schwefel, H. P. and Manner, R. (eds.), Springer (Berlin).
\item {\frenchspacing\sc De Menezes, M. A., Moukarzel, C. F., and Penna, T.
    J. P.}\ \ 2000\ \ First-order transition in small-world networks.
  Available as {\tt cond-mat/9903426}.
\item {\frenchspacing\sc Dorogovtsev, S. N. and Mendes, J. F. F.}\ \ 1999\ 
  \ Exactly solvable analogy of small-world networks.  Available as {\tt
  cond-mat/9907445}.
\refer{Erd\"os, P. and R\'enyi, A.}{1959}{On random graphs.}{\it
  Publicationes Mathematicae\/}{6}{290--297}
\refer{Fisher, M. E. and Berker, A. N.}{1982}{Scaling for first-order phase
  transitions in thermodynamic and finite systems.}{\it Physical Review
  B\/}{26}{2507--2513}
\item {\frenchspacing\sc Gladwell, M.}\ \ 1998\ \ Six degrees of Lois
  Weisberg.  {\it The New Yorker,} {\bf74}, No.~41, 52--64.
\bookref{Guare, J.}{1990}{Six Degrees of Separation: A Play.}{Vintage (New
  York)}
\item {\frenchspacing\sc Kasturirangan, R.}\ \ 1999\ \ Multiple scales in
  small-world graphs.  Massachusetts Institute of Technology AI Lab Memo
  1663.  Also {\tt cond-mat/9904055}.
\item {\frenchspacing\sc Kirby, D. and Sahre, P.}\ \ 1998\ \ Six degrees of
  Monica.  {\it New York Times}, February 21, 1998.
\refer{Kleczkowski, A. and Grenfell, B. T.}{1999}{Mean-field-type
  equations for spread of epidemics: The `small-world' model.}{\it
  Physica A\/}{274}{355--360}
\item {\frenchspacing\sc Kleinberg, J.}\ \ 1999\ \ The small-world
  phenomenon: An algorithmic perspective.  Cornell University Computer
  Science Department Technical Report 99--1776.  Also {\tt
    http://\-www.cs.cornell.edu/home/kleinber/\-swn.ps}.
\bookref{Kocken, M.}{1989}{The Small World.}{Ablex (Norwood, NJ)}
\refer{Korte, C. and Milgram, S.}{1970}{Acquaintance linking between white
  and negro populations: Application of the small world problem.}{\it
  Journal of Personality and Social Psychology\/}{15}{101--118}
\refer{Kretschmar, M. and Morris, M.}{1996}{Measures of concurrency in
networks and the spread of infectious disease.}{\it Mathematical
Biosciences\/}{133}{165--195}
\item {\frenchspacing\sc Kulkarni, R. V., Almaas, E., and Stroud, D.}\ \ 
  1999\ \ Evolutionary dynamics in the Bak-Sneppen model on small-world
  networks.  Available as {\tt cond-mat/9905066}.
\refer{Kulkarni, R. V., Almaas, E., and Stroud, D.}{2000}{Exact results
  and scaling properties of small-world networks.}{\it Physical Review
  E\/}{61}{4268--4271}
\refer{Lago-Fern\'andez, L. F., Huerta, R., Corbacho, F., and Sig\"uenza,
  J. A.}{2000}{Fast response and temporal coherent oscillations in
  small-world networks.}{\it Physical Review Letters\/}{84}{2758--2761}
\refer{Milgram, S.}{1967}{The small world problem.}{\it Psychology
  Today\/}{2}{60--67}
\refer{Monasson, R.}{1999}{Diffusion, localization and dispersion relations
  on small-world lattices.}{\it European Physical Journal
  B\/}{12}{555--567}
\refer{Moore, C. and Newman, M. E. J.}{2000a}{Epidemics and percolation in
  small-world networks.}{\it Physical Review E\/}{61}{5678--5682}
\item {\frenchspacing\sc Moore, C. and Newman, M. E. J.}\ \ 2000b\ \ Exact
  solution of site and bond percolation on small-world networks.  Available
  as {\tt cond-mat/0001393}.
\refer{Moukarzel, C. F.}{1999}{Spreading and shortest paths in systems
  with sparse long-range connections.}{\it Physical Review
  E\/}{60}{6263--6266}
\bookref{Newman, M. E. J. and Barkema, G. T.}{1999}{Monte Carlo Methods in
Statistical Physics.}{Oxford University Press (Oxford)}
\refer{Newman, M. E. J., Moore, C., and Watts, D. J.}{2000}{Mean-field
  solution of the small-world network model.}{\it Physical Review
  Letters\/}{84}{3201--3204}
\refer{Newman, M. E. J. and Watts, D. J.}{1999a}{Renormalization group
  analysis of the small-world network model.}{\it Physics Letters
  A\/}{263}{341--346}
\refer{Newman, M. E. J. and Watts, D. J.}{1999b}{Scaling and percolation
  in the small-world network model.}{\it Physical Review
  E\/}{60}{7332--7342}
\item {\frenchspacing\sc Remes, T.}\ \ 1997\ \ Six Degrees of Rogers
  Hornsby.  {\it New York Times}, August 17, 1997.
\refer{Sattenspiel, L. and Simon, C. P.}{1988}{The spread and persistence
of infectious diseases in structured populations.}{\it Mathematical
Biosciences\/}{90}{367--383}
\item {\frenchspacing\sc Tjaden, B. and Wasson, G.}\ \ 1997\ \ Available on
  the internet at {\tt http://www.cs.virginia.edu/oracle/}.
\item {\frenchspacing\sc Walsh, T.}\ \ 1999\ \ In {\it Proceedings of the
  16th International Joint Conference on Artificial Intelligence,}
  Stockholm, 1999.
\bookref{Watts, D. J.}{1999}{Small Worlds.}{Princeton University Press
  (Princeton)}
\refer{Watts, D. J. and Strogatz, S. H.}{1998}{Collective dynamics of
  ``small-world'' networks.}{\it Nature\/}{393}{440--442}
\end{list}

\end{document}